\documentclass{elsarticle}
\usepackage[a4paper, total={6in, 10in}]{geometry}
\usepackage{booktabs}
\usepackage{float}
\usepackage{amsmath}
\usepackage{amsthm}
\usepackage{amsmath,amssymb,amsfonts,mathrsfs,hyperref,microtype,amsthm}
\usepackage{graphicx}
\usepackage{graphicx,mathabx}
\usepackage{enumitem}  

\graphicspath{ {images/} }
\numberwithin{equation}{section}

\newtheorem{theorem}{Theorem}[section]
\newtheorem{proposition}{Proposition}[section]
\newtheorem{lemma}{Lemma}[section]

\newtheorem{example}{Example}
\newtheorem*{example*}{Example}
\newcommand{\BS}{\boldsymbol}

\newcommand{\Rmnum}[1]{\expandafter\@slowromancap\romannumeral #1@}

\journal{}
\makeatletter
\def\ps@pprintTitle{%
   \let\@oddhead\@empty
   \let\@evenhead\@empty
   \def\@oddfoot{\reset@font\hfil\thepage\hfil}
   \let\@evenfoot\@oddfoot
}

\begin{document}

\begin{frontmatter}
\author[a]{Helmi Shat\corref{cor1}}
\cortext[cor1]{Corresponding author}
\ead{hshat@ovgu.de}
\author[a]{Rainer Schwabe}
\ead{rainer.schwabe@ovgu.de}
\address[a]{\small Institute for Mathematical Stochastics, Otto-von-Guericke University Magdeburg, \\ \small Universit\"atsplatz 2,
39106 Magdeburg, Germany }

 \title{Optimal Time Plan in Accelerated Degradation Testing}

\begin{abstract}
Many highly reliable products are designed to function for years without failure. For such systems
accelerated degradation testing may provide significance information about the reliability properties of the system.
In this paper, we propose the $c$-optimality criterion for obtaining optimal designs of constant-stress accelerated degradation tests where the degradation path follow a linear mixed effects model. The present work is mainly concerned with developing optimal desings in terms of the time variable rather than considering an optimal design of the stress variable which is usually considered in the majority of the literature.
Finally, numerical examples are presented and sensitivity analysis procedures are conducted for evaluating the robustness of the optimal as well as standard designs against misspecifications of the nominal values.  
\end{abstract}

\begin{keyword}
Accelerated degradation test\sep linear mixed effects model\sep    $c$-optimal design\sep.

\end{keyword}

\end{frontmatter}

\section{Introduction}
\label{sec-introduction}
The growing research focus in recent years on the analysis of degradation
data has played a major role in the field of reliability inference and management. In particular, reliability data and characteristics
for highly reliable products can be collected through analyzing the obtained degradation
data. Consequently, statistical modeling and inference approaches have been introduced on the basis of
different degradation measuring techniques.

\cite{limon2017literature}, for instance, provided a comprehensive reviews on several approaches in the literature that are utilized to evaluate reliable information from acclerated degradation data.
\cite{RePEc:eee:reensy:v:142:y:2015:i:c:p:369-377} introduced an
analytical optimal ADT design prcedure for more efficient reliability demonstration by minimizing the
asymptotic variance of decision variable in reliability demonstration under the constraints of sample size, test duration, test cost, and predetermined decision risks.  Considering linear mixed effects model (LMEM), \citep{doi:10.1002/asmb.2061} considered the minimum asymptotic variance criterion to characterize optimal design as well as compromise design plans for accelerated degradation tests. Further, \cite{7114339} developed an algorithm-based optimal
ADT approach by minimizing the asymptotic variance
of the MLE of the mean failure time of a system, where
the sample size and termination time of each run of the ADT at a
constant measurement frequency were specified.
\citep{shat2021experimental} charachterized optimal designs for repeated measures accelerated degradation tests with multiple stress variables, where the degradation paths are assumed to follow a linear mixed effects. \citep{7160792} introduced a modeling precedure to
simultaneously analyze linear degradation data and traumatic
failures with competing risks in step stress ADT experiment.

The general theory of optimal design of experiments is efficiently charachterized in terms of the mathematical context of approximate designs (see e.\,g.\ \cite {silvey1980optimal}). For example, \cite{schwabe1996optimum} suggested optimal designs for multi-factor models which can be used to treat more than one stress variable and the choice of time plans simultaneously under various interaction structures. 

\cite{dette2010optimal} considered the problem of constructing $D$-optimal designs for linear and  nonlinear random effect models with applications in population
pharmacokinetics. 
%
\cite{grasshoff2012optimal} presented $D$-optimal designs for random coefficient regression models in the basis of geometrical arguments when only one observation is available per unit, i.e. a situation which occurs in destructive testing. 

It can be noted that the vast majority of the existing literature focuses on desinging the degradation stress variable(s). As opposed to this approach the time variable will be considered in the present work in order to obtain optimal experimental designs under various measurement plans.

The present paper is organized as follows. 
In Sections~\ref{sec-model-formulation} we state the general model formulation and in Section~\ref{sec-information} we exhibit the corresponding information matrix. Section~\ref{sec-failuretime} is devoted to derive the design optimality criterion for estimating a quantile of the failure time distribution under normal use conditions.
In Section~\ref{sec-opt-time-plan} also the measurement times are optimized under the constraint that measurements are taken according to the same time plan for all units, and in 
Section~\ref{sec-single-measurements} optimal measurement times are determined in the setting of destructive testing.
The paper closes with a short discussion in Section~\ref{sec-discussion}.

\section{Formulation of the model}
\label{sec-model-formulation}

In this section, we introduce a mixed effects regression model incorporating a product-type structure as in \citep{shat2021experimental}. 
In accordance with \citep{shat2021experimental} we assume that there are $n$ testing units $i=1,...,n$, for which degradation $y_{i j}$ will be measured according to a time plan $\mathbf{t}=(t_1,...,t_k)^T$ with $k$ time points $t_j$, $j = 1, ..., k$, $0 \leq t_1 < ... < t_k \leq t_{\max}$.
Each unit $i$ is measutred under a setting $\mathbf{x}_i$ of the stress variable(s), which is kept fixed for each unit throughout the testing, but varies between units.
The number $k$ of measurements and the measurent times $t_1,...,t_k$ are the same for all units.  
The measurements $y_{i j}$ are considered as realizations of random variables $Y_{i j}$ which can be charachterized by a hierarchical model. 
For each unit $i$ the observation $Y_{i j}$ at $t_j$ is expressed as
\begin{equation} 
\label{modelindividualresponse}
Y_{i j} = \mu_{i}(\mathbf{x}_i, t_j) + \varepsilon_{i j} ,
\end{equation}
where $\mu_{i}(\mathbf{x}, t)$ is the mean degradation of unit $i$ at time $t$, when stress $\mathbf{x}$ is applied, and $\varepsilon_{i j}$ is a measurement error.
The mean degradation $\mu_{i}(\mathbf{x}, t)$ is given by a linear model equation in the stress variable $\mathbf{x}$ and the time $t$,
\begin{equation} 
	\label{mean-degradation-unit}
	\mu_{i}(\mathbf{x}, t) = \sum_{q = 1}^{p} \beta_{i q} f_{q}(\mathbf{x}, t_j) 
	= \mathbf{f}(\mathbf{x}, t_j)^{T} \BS{\beta}_i  
\end{equation}
where $\mathbf{f}(\mathbf{x}, t) = (f_{1}(\mathbf{x}, t), ..., f_{p}(\mathbf{x}, t))^T$ is a $p$-dimensional vector of known regression functions $f_{q}(\mathbf{x}, t)$ in both the stress variable(s) $\mathbf{x}$ and the time $t$, and $\BS{\beta}_i = (\beta_{i, 1}, ..., \beta_{i, p})^T$ is a $p$-dimensional vector of unit specific parameters $\beta_{i, q}$.
Thus, the response is given by
\begin{equation} 
	\label{modelindividuallevel}
	Y_{i j} = \mathbf{f}(\mathbf{x}_{i}, t)^{T}\BS{\beta}_i + \varepsilon_{i j}.
\end{equation}
Further, the measurement error $\varepsilon_{i j}$ is assumed to be normally distributed with zero mean and some potentially time dependent error variance $\sigma_{\epsilon j}^{2}$. 
The vector $\BS{\varepsilon}_{i} = (\varepsilon_{i 1}, ..., \varepsilon_{i k})^T$ of errors within one unit $i$ is $k$-dimensional multivariate normal with zero mean and positive definite variance covariance matrix $\BS{\Sigma}_\varepsilon$. 
For the regression functions $\mathbf{f}(\mathbf{x},t)$ we suppose a product-type structure with complete interactions between the stress variable $\mathbf{x}$ and the time $t$, i.\,e.\ there are marginal regression functions $\mathbf{f}_1(\mathbf{x})=(f_{1 1}(\mathbf{x}), ..., f_{1 p_1}(\mathbf{x}))^T$ and $\mathbf{f}_2(t)=(f_{2 1}(t), ..., f_{2 p_2}(t))^T$ of dimension $p_1$ and $p_2$ which only depend on the stress variable $\mathbf{x}$ and the time $t$, respectively, such that the vector $\mathbf{f}(\mathbf{x}, t) = \mathbf{f}_1(\mathbf{x}) \otimes \mathbf{f}_2(t)$ of regression functions factorizes into its marginal counterparts ($p = p_1 p_2$).
Here ``$\otimes$'' denotes the Kronecker product.
Then the observation $Y_{i j}$ can be written as
\begin{equation} 
	\label{modelindividuallevel-product}
	Y_{i j} = \sum_{r = 1}^{p_1} \sum_{s = 1}^{p_2} \beta_{i rs} f_{1 r}(\mathbf{x}_{i}) f_{2 s}(t_j) + \varepsilon_{i j}
	= (\mathbf{f}_1(\mathbf{x}_i) \otimes \mathbf{f}_2(t_j))^T \BS{\beta}_i + \varepsilon_{i j},
\end{equation}
where the entries of the vector $\BS{\beta}_i = (\beta_{i 1 1}, ..., \beta_{i 1 p_2}, ..., \beta_{i p_1 p_2})$ of parameters are relabeled lexicographically according to the marginal regression functions.
Moreover, 
we will assume throughout that the marginal regression function $\mathbf{f}_1(\mathbf{x}) = (f_{11}(\mathbf{x}), ..., f_{1 p_1}(\mathbf{x}))^T$ of the stress variable $\mathbf{x}$ contains a constant term, $f_{1 1}(\mathbf{x}) \equiv 1$ say, 
and that only the leading $p_2$ parameters $\beta_{i 1 1}, ..., \beta_{i 1 p_2}$ of $\BS{\beta}_i$ associated with this constant term are unit specific. 
The remaining parameters in $\BS{\beta}_i$ are assumed to have the same value $\beta_{irs}=\beta_{rs}$, $r = 2, ..., p_1$, $s = 1, ..., p_2$, for all units $i = 1, ..., n$.
Hence, for unit $i$ the model \eqref{modelindividuallevel-product} can be rewritten as
\begin{equation} 
\label{eq-general-model}
 Y_{i j} = \left(\mathbf{f}_1(\mathbf{x}_{i}) \otimes \mathbf{f}_2(t_j)\right)^{T}\BS{\beta} + \mathbf{f}_2(t_j)^T\BS{\gamma}_i + \varepsilon_{i j},
\end{equation}
where $\BS{\beta} = (\beta_{1 1}, ..., \beta_{p_1 p_2})^T$ is the vector of (aggregate) fixed effect parameters (averaged over the units) 
 and $\BS{\gamma}_i = (\gamma_{i 1}, ..., \gamma_{i p_2})^T$ is the $p_2$-dimensional vector of unit specific deviations $\gamma_{i s} = \beta_{i, 1 s} - \beta_{1 s}$, $s = 1, ..., p_2$, from the corresponding aggregate parameters where $\BS\gamma_i$ are independent and identically distributed multivariate normal with mean 0 and covariance matrix $\BS\Sigma_\gamma$.
For the analysis of degradation under normal use, we further assume that the general model~\ref{eq-general-model} is also valid at the normal use condition $\mathbf{x}_u$, i.\,e.\ 
\begin{equation}
	\label{eq-degr-path-use-cond}
	\mu_u(\mathbf{x}_{u},t) = (\mathbf{f}_1(\mathbf{x}_{u}) \otimes \mathbf{f}_2(t))^{T} \BS{\beta} + \mathbf{f}_2(t)^T \BS{\gamma}_u
\end{equation} 
describes the mean degradation of a future unit $u$ at normal use condition $\mathbf{x}_u$ and time $t$, and the random effects $\BS{\gamma}_u$ are multivariate normal with mean zero and variance covariance matrix $\BS\Sigma_\gamma$.

\section{Information and design}  
\label{sec-information}
Denote by $\BS\theta = (\BS{\beta}^T, \BS{\varsigma}^T)^T$ the vector of all model parameters, where $\BS\varsigma$ collects all variance covariance parameters from $\BS\Sigma_\gamma$ and $\BS\Sigma_\varepsilon$. Let $ \mathbf{F}_1 = (\mathbf{f}_1(\mathbf{x}_1), ..., \mathbf{f}_1(\mathbf{x}_n))^{T}$ be the $n \times p_1$ marginal design matrix for the stress variables across units, and denote by $\mathbf{F}_2 = ( \mathbf{f}_2(t_1), ..., \mathbf{f}_2(t_k))^T$ the $k \times p_2$ marginal design matrix for the time variable. The corresponding per unit variance covariance matrix is $\mathbf{V}=\mathbf{F}_2{\BS{\Sigma}}_\gamma\mathbf{F}_2^T + {\BS{\Sigma}}_{\varepsilon}$ 
where $\mathbf{V}$ depends only on $\BS\varsigma$, see \citep{shat2021experimental} for further details about estimation in the model parameters.
The Fisher information matrix is block diagonal,
\begin{equation}
	\label{info-block}
	\mathbf{M}_{\BS{\theta}}
	= \left( 
		\begin{array}{cc}
			\mathbf{M}_{\BS\beta} & \mathbf{0}
			\\
			\mathbf{0} & \mathbf{M}_{\BS\varsigma}
		\end{array}
	\right) 
\end{equation}  
where $\mathbf{M}_{\BS\beta}$ and $\mathbf{M}_{\BS\varsigma}$ are the information matrices for $\BS\beta$ and $\BS\varsigma$, respectively, for short.
The particular form of $\mathbf{M}_{\BS\varsigma}$ will be not of interest. 
 By (\ref{eq-degr-path-use-cond}), the information matrix $\mathbf{M}_{\BS\beta}$ for the parameters $\BS\beta$ factorizes according to
\begin{equation}
	\label{eq-info-product}
	\mathbf{M}_{\BS\beta} = \mathbf{M}_1 \otimes \mathbf{M}_2
\end{equation}
into the information matrix $\mathbf{M}_1 = \mathbf{F}_1^T \mathbf{F}_1$ in the marginal model 
in the stress variable and the information matrix $\mathbf{M}_2 = \mathbf{F}_2^T\mathbf{V}^{-1}\mathbf{F}_2$ in the marginal mixed effects model in the time variable $t$ where the time plan $\BS t =(t_1,...,t_k)^T$ plays the role of a (exact) design here.
\section{Optimality criterion based on the failure time under normal use condition}
\label{sec-failuretime}
In accordance with \cite{doi:10.1002/asmb.2061} we are focusing on some characteristics of the failure time distribution of soft failure due to degradation. 
As already mentioned, we assume that the model equation (\ref{eq-degr-path-use-cond}) $\mu_{u}(t) = \mu(\mathbf{x}_{u}, t) = (\mathbf{f}_1(\mathbf{x}_{u}) \otimes \mathbf{f}_2(t))^{T} \BS{\beta} + \mathbf{f}_2(t)^T \BS{\gamma}_u$ for the mean degradation paths is also valid under normal use condition $\mathbf{x}_u$.
A soft failure due to degradation is expressed as the exceedance of the mean degradation $\mu_u$ over a  failure threshold $y_0$. The failure time $T$ under normal use condition is defined as the first time $t$ the mean degradation path $\mu_u(t)$ reaches or exceeds the threshold $y_0$, i.\,e.\ $T = \min \{t \geq 0;\, \mu_u(t) \geq y_0\}$.
As random effects $\BS\gamma_u$ are involved in the mean degradation path, the failure time $T$ is random. As in \citep{shat2021experimental} we denote by $\mu(t) = \mathrm{E}(\mu_u(t)) = (\mathbf{f}_1(\mathbf{x}_{u}) \otimes \mathbf{f}_2(t))^{T}\BS{\beta}$ the aggregate degradation path under normal use condition and by $\BS\delta  = (\delta_1, ..., \delta_{p_2})^T$ the vector of its coefficients $\delta_{s} = \sum_{r = 1}^{p_1} f_{1 r}(\mathbf{x}_u)\beta_{r s}$, in the regression functions $f_{2 s}$ in $t$, i.\,e.\ $\mu(t) = \mathbf{f}_2(t)^T\BS\delta = \sum_{s = 1}^{p_2}\delta_s f_{2 s}(t)$.\\

For the purpose of describing certain characteristics of the distribution of the failure time, we will determine the distribution function $F_T(t) = \mathrm{P}(T \leq t)$.
First note that $T \leq t$ if and only if $\mu_u(t) \geq y_0$.
Hence
\begin{eqnarray}
   F_{T}(t) &  = & \mathrm{P}(\mu_u(t) \geq y_0)
   \nonumber
	\\
	& = & \mathrm{P} (- \mathbf{f}_2(t)^T \BS{\gamma}_u \leq \mu(t) - y_0) 
   \nonumber
   \\
	& = & \Phi(h(t)) ,
    \label{eq-failure-time-distribution}
\end{eqnarray}
where
\begin{equation}
	\label{eq-h-tau}
	h(t) = \frac{\mu(t) - y_0}{\sigma_u(t)} ,
\end{equation}
$\sigma_u^2(t) = \mathbf{f}_2(t)^T \BS\Sigma_\gamma \mathbf{f}_2(t)$ is the variance of the mean degradation path $\mu_u(t)$ at time $t$, and $\Phi$ denotes the standard normal distribution function.

We will be interested in quantiles $t_\alpha$ of the failure time distribution, $\mathrm{P}(T \leq t_\alpha) = \alpha$, up to which under normal use conditions $\alpha \cdot 100$ percent of the units fail and $(1 - \alpha) \cdot 100$ percent of the units persist.
Of particular importance is the median $t_{0.5}$.
By (\ref{eq-failure-time-distribution}) these quantiles can be determined as the solutions of the equation $h(t_\alpha) = z_\alpha$, where $z_\alpha = \Phi^{-1}(\alpha)$ is the $\alpha$-quantile of the standard normal distribution.
For the median ($\alpha = 0.5$) we have $z_{0.5} = 0$ and, hence the median failure time $t_{0.5}$ is the solution of $\mu(t) = y_0$. 
In the particular case of straight lines for the mean degradation paths, i.\,e.\ $\mathbf{f}_2(t) = (1,t)^T$ and $\BS\Sigma_\gamma =\left(\begin{array}{cc}
	\sigma_1^2 & \rho \sigma_1 \sigma_2
	\\
	\rho \sigma_1 \sigma_2 & \sigma_2^2
\end{array}\right)$
where $\sigma_1^2, \sigma_2^2$ and $\rho$ are defined, 
the function $h(t)$ specifies to
\begin{equation}
	\label{eq-h-tau-linear}
	h(t) = \frac{\delta_2 t + \delta_1 - y_0}{\sqrt{\sigma_1^2 + 2\rho \sigma_1 \sigma_2 t + \sigma_2^2 t^2}} ,
\end{equation}
where $\delta_{1} = \sum_{r = 1}^{p_1} f_{1 r}(\mathbf{x}_u) \beta_{r 1}$ and $\delta_2 = \sum_{r = 1}^{p_1} f_{1 r}(\mathbf{x}_u) \beta_{r 2}$ are the intercept and the slope of the aggregate degradation path $\mu(t) = \delta_1 + \delta_2 t$ under normal use condition, respectively.
The median failure time is then given by $t_{0.5} = (y_0 - \delta_1) / \delta_2$ which gives a solution $t_{0.5} > 0$ under the natural assumptions that the aggregate degradation path is increasing, $\delta_2 > 0$, and that the aggregate degradation at time $t=0$ is less than the failure threshold, $\delta_1 < y_0$.

In general, under the assumption that the correlation of the random effects is non-negative for the intercept and the slope of the mean degradation path,  $\rho \geq 0$, the function $h(t)$ can be seen to be strictly increasing, $h^\prime(t) > 0$, in $t>0$.
Note that $\Phi(h(0)) = \Phi(-(y_0-\delta_1)/\sigma_1)$ is the probability that under normal use condition the mean degradation path exceeds the threshold $y_0$ already at the initial time $t=0$ where $\lim_{t\to\infty}h(t) = \delta_2/\sigma_2$
and $t_\alpha$ is nondegenerate for
$\Phi(-(y_0-\delta_1)/\sigma_1) < \alpha < \Phi(\delta_2/\sigma_2)$.
In the special case of only a random intercept in the random effects, i.\,e.\ $\sigma_2^2 = 0$, all $\alpha$-quantiles $t_\alpha$ finitely exist for $\alpha \geq \Phi( - (y_0-\delta_1) / \sigma_1)$ and can be determined as $t_\alpha = (y_0 - \delta_1 + z_\alpha\sigma_1) / \delta_2$.

The quantile $t_\alpha = t_\alpha(\BS\theta)$ is a function of both the  fixed effect parameters $\BS\beta$ and the variance parameters $\BS\varsigma$, in general.

By the delta-method the maximum likelihood estimator $\widehat{t}_\alpha$ of ${t}_\alpha$ is asymptotically normal with  asymptotic variance
\begin{equation}
	\label{eq-avar-tau-alpha}
	\mathrm{aVar}(\widehat{t}_{\alpha}) = \mathbf{c}^T \mathbf{M}_{\BS{\theta}}^{-1} \mathbf{c} ,
\end{equation}
where $\mathbf{c} = \frac{\partial}{\partial\BS\theta}t_{\alpha}$ is the gradient vector of $t_{\alpha}$ with respect to the parameter vector ${\BS\theta}$. 
The asymptotic variance depends on the design of the experiment through the information matrix $ \mathbf{M}_{\BS{\theta}}$ and will be chosen as the optimality criterion for the design.
The gradient $\mathbf{c}$ is given by
\begin{equation}
	\label{eq-gradient-tau-alpha}
	\mathbf{c} =  c_0 \left({\textstyle{\frac{\partial}{\partial\BS\theta}}}\mu(t)\vert_{t=t_\alpha} - z_\alpha {\textstyle{\frac{\partial}{\partial\BS\theta}}}\sigma_u(t)\vert_{t=t_\alpha}\right) ,
\end{equation}
in view of (\ref{eq-h-tau}) by the implicit function theorem (see e.\,g.\ \cite{krantz2012implicit}), where   $c_0 \neq 0$ is a constant not dprending on the design.
As the aggregate mean degradation $\mu(t)$ only depends on the aggregate location parameters $\BS\beta$ and the variance $\sigma_u^2(t)$ only depends on the variance parameters $\BS\varsigma$ the gradient simplifies to $\mathbf{c} = - c_0 (\mathbf{c}_{\BS\beta}^T, \mathbf{c}_{\BS\varsigma}^T)^T$, where
\begin{equation*}
	\label{eq-gradient-tau-alpha-beta}
	\mathbf{c}_{\BS\beta} = {\textstyle{\frac{\partial}{\partial\BS\beta}}}\mu(t)\vert_{t = t_\alpha} = \mathbf{f}(\mathbf{x}_u, t_\alpha)
\end{equation*}
is the gradient of $\mu(t)$ with respect to $\BS\beta$ and
\begin{equation*}
	\label{eq-gradient-tau-alpha-sigma}
	\mathbf{c}_{\BS\varsigma} = - z_\alpha {\textstyle{\frac{\partial}{\partial\BS\varsigma}}}\sigma_u(t)\vert_{t = t_\alpha}
\end{equation*}
is, up to a multiplicative constant, the gradient of $\sigma_u(t)$ with respect to $\BS\varsigma$.
The particular shape of $\mathbf{c}_{\BS\varsigma}$ does not play a role here, in general.
But it is important that $\mathbf{c}_{\BS\varsigma} = \mathbf{0}$ in the case of the median ($\alpha=0.5$).
By the block diagonal form (\ref{info-block}) of the information matrix the asymptotic variance (\ref{eq-avar-tau-alpha}) of $\widehat{t}_\alpha$ becomes
\begin{equation}
	\label{eq-avar-tau-alpha-sum}
	\mathrm{aVar}(\widehat{t}_{\alpha}) = c_0^2 \left(\mathbf{f}(\mathbf{x}_u, t_\alpha)^T \mathbf{M}_{\BS{\beta}}^{-1}\mathbf{f}(\mathbf{x}_u, t_\alpha) +z_\alpha^2 \frac{\partial}{\partial\BS\varsigma}\sigma_u(t_\alpha)^T \mathbf M_{\BS \varsigma}^{-1} \frac{\partial}{\partial\BS\varsigma}\sigma_u(t_\alpha)\right) 
\end{equation}
which simplifies to $\mathrm{aVar}(\widehat{t}_{0.5}) = c_0^2 \mathbf{f}(\mathbf{x}_u, t_\alpha)^T \mathbf{M}_{\BS{\beta}}^{-1} \mathbf{f}(\mathbf{x}_u, t_\alpha)$
in the case of the median. For the case of a product-type model the expression related to the fixed effect parameters $\BS\beta$ decomposes into
\begin{equation}
	\label{eq-avar-tau-alpha-decompose}
	\mathbf{c}_{\BS\beta}^T \mathbf{M}_{\BS{\beta}}^{-1} \mathbf{c}_{\BS\beta} = \mathbf{f}_1(\mathbf{x}_u)^T \mathbf{M}_{1}^{-1} \mathbf{f}_1(\mathbf{x}_u) \cdot \mathbf{f}_2(t_\alpha)^T \mathbf{M}_{2}^{-1} \mathbf{f}_2(t_\alpha).
\end{equation}

\section{Optimization of the time plan}
\label{sec-opt-time-plan}

In  \citep{shat2021experimental} we treated the situation when there is a fixed time plan $\mathbf{t} = (t_1, ..., t_k)^T$, and only the stress variables $\mathbf{x} \in \mathcal X$ are to be optimized across units. There we noticed that the optimal design for the stress variable does not depend on the time plan used.
We consider now the situation when also the settings $t_1, ..., t_k$ for the standardized time variable $t \in [0, 1]$ may be optimized within units.
As in Section~\ref{sec-model-formulation} we still assume that the same time plan  $\mathbf{t}=(t_1, ..., t_k)^T$ is used for all units. By equations (\ref{eq-avar-tau-alpha-sum}) and (\ref{eq-avar-tau-alpha-decompose}), the asymptotic variance is given by
\begin{equation}
	\label{eq-avar-tau-alpha-sum-decompose}
	\mathrm{aVar}(\widehat{t}_{\alpha}) = c_0^2 \left(\mathbf{f}_1(\mathbf{x}_u)^T \mathbf{M}_{1}^{-1} \mathbf{f}_1(\mathbf{x}_u) \cdot \mathbf{f}_2(t_\alpha)^T \mathbf{M}_{2}^{-1} \mathbf{f}_2(t_\alpha) + \mathbf{c}_{\BS\varsigma}^T {\mathbf{M}}_{\BS{\varsigma}}^{-1} \mathbf{c}_{\BS\varsigma}\right) 
\end{equation}

Here the dependence of the asymptotic variance on the settings at the time plan $\mathbf{t}$ for the time variable comes in through the information matrix $\mathbf{M}_{2}$ in the second marginal model as well as through the information matrix ${\mathbf{M}}_{\BS{\varsigma}}$ for the variance parameters.
In general, the asymptotic variance is a compound criterion in $\mathbf{f}$ which aims at minimizing a linear combination of a variance term $\mathbf f_2(t_\alpha)^T \mathbf M_2 ^{-1} \mathbf f_2(t_\alpha)$ in the location parameters $\BS\beta$ and a variance term $\mathbf c_{\BS\varsigma}^T \mathbf M_{\BS\varsigma}^{-1} \mathbf c_{\BS\varsigma}$ in the variance parameters. In this compound criterion the first coefficient involves the marginal information matrix $\mathbf M_1$ for the stress variable. As a consequence the criterion and, hence, the optimal time plan depend on the design for the stress variable. Thus first the marginal desiogn for the stress variable has to be optimized which can be done simultaneously in all time plans. Then the corresponding value has to be inserted into the compound criterion for optimizing the time plan. To facilitate further calculations we restrict to the case of estimating the median failure time $t_{0.5}$.
Consequently, the asymptotic variance simplifies to
\begin{equation}
	\label{eq-avar-tau-median-decompose}
	\mathrm{aVar}(\widehat{t}_{0.5}) = c_0^2 \mathbf{f}_1(\mathbf{x}_u)^T \mathbf{M}_{1}^{-1} \mathbf{f}_1(\mathbf{x}_u) \cdot \mathbf{f}_2(t_{0.5})^T \mathbf{M}_{2}^{-1} \mathbf{f}_2(t_{0.5}) \, .
\end{equation}
Hence, for the median failure time also the optimization of the measurement times $t_1,..,t_k$ can be performed independently of the marginal model of the stress variables and their settings.
Only the marginal $c$-criterion $\mathbf{f}_2(t_{0.5})^T \mathbf{M}_{2}^{-1} \mathbf{f}_2(t_{0.5})$ has to be minimized. Remember that the marginal information matrix $\mathbf{M}_{2}$ is given by $\mathbf{M}_{2} = \mathbf{F}_{2}^T \mathbf{V}^{-1} \mathbf{F}_{2}$ where the variance covariance matrix $\mathbf{V}$  of the vector $\mathbf{Y}_i$ of measurements for each unit $i$ is given by $\mathbf{V} = \mathbf{F}_2 {\BS{\Sigma}}_\gamma \mathbf{F}_2^T + {\BS{\Sigma}}_{\varepsilon}$.
In the situation of uncorrelated and homoscedastic measurment erros ($\BS{\Sigma}_\varepsilon = \sigma^2_\varepsilon\mathbf{I}_k$), \citep{Schmelter2007} provided a representation of the invere information matrix in random effects models.
This can be readily extended through the following result to a general, non-singular variance covariance structure $\BS{\Sigma}_\varepsilon$ for the measurement errors  $\BS\varepsilon$.
\begin{lemma}
	\label{lem-repr-minv}
	Let $\mathbf{F}$ be a $k \times p$ matrix of rank $p$, $\BS\Sigma_\gamma$  a non-negative definite $p \times p$ matrix, $\BS\Sigma_\varepsilon$ a positive definite $k \times k$ matrix, $\mathbf{V} = \mathbf{F} \BS\Sigma_\gamma \mathbf{F}^T + \BS\Sigma_\varepsilon$, and $\mathbf{M} = \mathbf{F}^T \mathbf{V}^{-1} \mathbf{F} $.
	Then  
	\begin{equation} 
		\mathbf{M}^{-1} = (\mathbf{F}^T \BS\Sigma_\varepsilon^{-1} \mathbf{F})^{-1} + \BS\Sigma_\gamma .
	\end{equation}  
\end{lemma}
where the proof of Lemma~\ref{lem-repr-minv} is obtained in the Appendix.
From this we obtain for the marginal information matrix $\mathbf{M}_{2}$ that its inverse can be decomposed to $\mathbf{M}_{2}^{-1} = \mathbf{F}_2^T \BS{\Sigma}_\varepsilon^{-1} \mathbf{F}_2+ \BS{\Sigma}_\gamma$. As a consequence, the marginal $c$-criterion $\mathbf{f}_2(t_{0.5})^T \mathbf{M}_{2}^{-1} \mathbf{f}_2(t_{0.5})$ can be split up into
\begin{equation}
	\label{eq-var-t-extra-decompose}
	\mathbf{f}_2(t_{0.5})^T \mathbf{M}_{2}^{-1} \mathbf{f}_2(t_{0.5}) = \mathbf{f}_2(t_{0.5})^T {\mathbf{M}_{2}^{(0)}}^{-1} \mathbf{f}_2(t_{0.5}) + \mathbf{f}_2(t_{0.5})^T \BS{\Sigma}_\gamma \mathbf{f}_2(t_{0.5}) ,
\end{equation}
where $\mathbf{M}_{2}^{(0)} = \mathbf{F}_2^T \BS{\Sigma}_\varepsilon^{-1} \mathbf{F}_2$ is the information matrix in the marginal fixed effect model in the time variable $t$.
Hence, as for any linear criterion, the optimization of the asymptotic variance~(\ref{eq-avar-tau-median-decompose}) of the median failure time ${t}_{0.5}$ with respect to the time plan $\mathbf t$ does not depend on the variance covariance matrix $\BS{\Sigma}_\gamma$ of the random effects.
For optimization only the term $\mathbf{f}_2(t_{0.5})^T {\mathbf{M}_{2}^{(0)}}^{-1} \mathbf{f}_2(t_{0.5})$ has to be minimized which is the $c$-criterion for extrapolation of the mean response at $t_{0.5}$ in the marginal fixed effect model in $t$.
This leads to the following result for optimization with respect to the time variable.

\begin{proposition}
	\label{prop-extrapolation-t}
	If the time plan $\mathbf t$ is $c$-optimal for extrapolation of the mean response at  $t_{0.5}$ in the marginal fixed effect model with covariance $\BS\Sigma_\varepsilon$ for the time variable $t$, then $\mathbf t$ minimize the asymptotic variance for the estimator $\widehat{t}_{0.5}$ of the median failure time $t_{0.5}$ under normal use condition.
\end{proposition}

Note that in degradation experiments the median failure time $t_{0.5}$ under normal use condition is typically much larger than the time horizon of the experiment, i.\,e.\ $t_{0.5} > 1$ on the standardized time scale.
However, the above proposition also holds for interpolation, $t_{0.5} \in [0,1]$. The optimal time plan depends on the location parameters $\BS\beta$ through $t_{0.5}$, but also on the variance covariance structure $\BS\Sigma_{\varepsilon}$ of the measurement errors within units, and is, hence, local.
If the number $k$ of measurement times $t_1, ..., t_k$ is large, one might be tempted to use the concept of approximate designs also here.

In that case, the approximate design $\tau$, as introduced by\citep{kiefer1959optimum}, will be defined by mutually distinct time points $t_1, ..., t_\ell$ from the standardized experimental time interval $\mathcal{T} = [0,1]$ with corresponding proportions $\pi_1, ..., \pi_\ell > 0$ satisfying $\sum_{j=1}^\ell \pi_j = 1$ with the interpretation that (approximately) $\pi_j k$ measurements are performed at time point $t_j$, $j = 1, ..., \ell$, for each unit.
In the situation of a linear time trend, $\mathbf{f}_2(t) = (1,t)^T$ this leads to essentially the same extrapolation problem as for the stress variable in \citep{shat2021experimental}.  Under the additional assumptions of uncorrelated and homoscedastic measurement errors
and $t_{0.5}>=1$, the $c$-optimal design for extrapolation at $t_{0.5} > 1$ is concentrated on the $\ell = 2$ endpoints $t_1 = 0$ and $t_2 = 1$ of the standardized experimental time interval $\mathcal{T}=[0,1]$ with corresponding proportions $\pi_1 = \frac{t_{0.5} - 1}{2 t_{0.5} - 1}$ and  $\pi_2 = \frac{t_{0.5}}{2 t_{0.5} - 1}$ (see \citep{schwabe1996optimum}, Example~2.1).
Similarly, as for extrapolation of the stress variable in \citep{shat2021experimental}, the proportion $\pi_2$ at the endpoint $t_2 = 1$ in the direction of the extrapolation time $t_{0.5}$ decreases from $1$, when $t_{0.5}$ is close to $1$, to $0.5$, when the distance gets large ($t_{0.5} \to \infty$).

However, from a practical point of view, it often does not seem meaningful to have replications, i.\,e.\ to have more than one measurement at a time at the same unit.
Moreover, even if this would be possible, these measurements would be expected to be strongly correlated with a correlation beyond that caused by the random effects. 
For instance, the different measurement times $t_1, ..., t_k$ may be assumed to be at least some sufficiently large $\Delta t > 0$ apart, or more specifically, they are restricted to a grid on the standardized experimental time interval with grid size $\Delta t$. 
Nevertheless approximate design theory can be used to determine optimal time plans.
To do so the standardized experimental time interval is discretized to a sufficiently coarse grid, $\mathcal{T} = \{j \Delta t;\, j = 0,1, ..., J\}$, where $\Delta t = 1/J$, i.\,e.\ $\mathcal{T} = \{0, \Delta t, 2 \Delta t, ..., 1\}$.
Additionally constraints are imposed on the proportions $\pi_j$ that none of these proportions is larger than $1/k$, i.\,e.\ the number of measurements at a time is bounded by one and, hence, there are at least $k$ different time points.
Optimal approximate time plans can then be obtained under these constraints by using standard algorithms for design optimization.
Actually, the so obtained proportions may be smaller than one for some of the time points. 
But a theoretical result based on an equivalence theorem under constraints (convex optimization with Kuhn-Tucker conditions) guarantees that this only occurs for a small number of time points.  For the $D$-criterion 
\citep{sahm2001note} showed that the experimental region splits up in subregions where the weight of the optimal design attains the upper bound $1/k$ and where it is $0$.
Only at the boundary of these subregions weights strictly between $0$ and $1/k$ may occur.
In particular, for straight line regression on $[0,1]$, the subregions with maximal weight $1/k$ are adjacent to the endpoints $0$ and $1$ of the interval while in the interior the optimal design has zero weight.
This result carries over directly to other criteria like the $c$-criterion under consideration.
From this approach efficient exact designs can be obtained by adjusting the weights to the admissible values $0$ and $1 / k$ under the constraint of total weight $1$. For brevity we consider in the following Example the model of Example~{2} in \citep{shat2021experimental} for numerical results.
\begin{table}
	\begin{center}
		\caption{Nominal values for Example~\ref{ex-2}}
		\label{tab-ex-intro-nominal-values}
		\vspace{2mm}
		\begin{tabular}{c|c|c|c|c||c|c|c|c||c||c}
			& $\beta_{1}$ & $\beta_2$ & $\beta_3$ & $\beta_4$ & $\sigma_{1}$ & $\sigma_{2}$ & $\rho$ & $\sigma_\varepsilon$ & ${x}_{u}$ & $y_0$
			\\
			$f_j(x,t)$ & $1$ & $x$ & $t$ & $x t$ & & & & & &
			\\
			\hline
			& $2.397$ & $1.629$ & $1.018$ & $0.0696$ & $0.114$ & $0.105$ & $-0.143$ & $0.048$ & $-0.056$ & $3.912$ 		
		\end{tabular} 
	\end{center}
\end{table} 
\begin{example}
\label{ex-2} In this model we have straight line regression in both a single stress variable $x$ and the time variable $t$ with an interaction term  $x t$. The response $y_{i j}$ of testing unit $i$ at time $t_j$ is given by
\begin{equation} 
\label{example2_111}
 y_{i j} = \beta_{i1} + \beta_2 x_{i } + \beta_{i 3} t_j + \beta_4 x_it_j + \varepsilon_{i j}
\end{equation}
under the nominal values of Table~\ref{tab-ex-intro-nominal-values} which are reproduced from Example~7.2 in \cite{doi:10.1002/asmb.2061} after standardization. The measurement errors are supposed to be uncorrelated and homoscedastic ($\BS\Sigma_{\varepsilon} = \sigma^2_{\varepsilon} \mathbf{I}_k$), and the measurement errors $\varepsilon_{i j}$ are realizations of a normally distributed error variable with mean zero and error variance $\sigma^2_\varepsilon$. Further, the unit specific parameters $(\beta_{i 1}, \beta_{i 3})^T$ of the units are assumed to be realizations of a bivariate normal distribution with mean $(\beta_{1}, \beta_{3})^T$ and a variance covariance matrix $\BS{\Sigma}_\gamma =
\left(\begin{array}{cc}
	\sigma_1^2 & \rho \sigma_1 \sigma_2
	\\
	\rho \sigma_1 \sigma_2 & \sigma_2^2
\end{array}\right)$.
	We are searching for a locally $c$-optimal design $\tau^*$ for extrapolation at the median failure time $t_{0.5}=1.583$.
	As constraints for the design we assume that $k = 6$ observation can be taken on a grid with increment $\Delta t = 0.05$ of the standardized experimental time interval $[0,1]$, i.\,e.\ $J = 20$ and $\mathcal{T} = \{0, 0.05, 0.10, ..., 1\}$. By the multiplicative algorithm (see e.g. \citep{silvey1978algorithm} ) adapted to the present constraint situation, the following numerical solution is obtained for the locally $c$-optimal design
	\[	
	\tau^* = 
		\left(
			\begin{array}{ccccccc}
				0.00 & 0.05 & 0.10 & 0.85 & 0.90 & 0.95 & 1.00 
				\\ 
				0.166 & 0.166 & 0.130 & 0.055 & 0.166 & 0.166 & 0.166
			\end{array} 
		\right) \, .
	\]	
	The optimal approximate design $\tau^*$ is supported on seven time points which are concentrated to the ends of the standardized experimental time interval $[0,1]$ with maximal admitted weight $1/k$ for all but the two boundary points $t = 0.10$ and $t  = 0.85$ separating the grid points $t = 0.00, 0.05, 0.90,  0.95$, and $ 1.00$ with full weight $1/k$ from those with zero weight $ (t=0.15,...,0.80).$
	This shape of the optimal design is in accordance with the findings of \citep{sahm2001note} for $D$-optimality.
	In view of Proposition~\ref{prop-extrapolation-t} the design $\tau^*$ is also optimal for the estimation of the median failure time.

	For practical use, the optimal approximate design $\tau^*$ may be adjusted to 
	\[
		\tau_0 = 
			\left(
				\begin{array}{cccccc}
					0.00 & 0.05 & 0.10 & 0.90 & 0.95 & 1.00 
					\\ 
					0.166 & 0.166 & 0.166 & 0.166 & 0.166 & 0.166
				\end{array} 
			\right)
	\]
	which is supported on exactly $k = 6$ time points with equal weights $1/k=1/6$ by transferring the weight from the boundary point $t= 0.85$ to the boundary point $t = 0.10$  (see \citep{dorfleitner1999rounding}). The design $\tau_0$ can be realized as an exact design by taking one measurement at each of the six time points $0.00$, $0.05$, $0.10$, $0.90$, $0.95$, and $1.00$.
	To quantify what might have got lost, the quality of the adjusted design $\tau_0$ can be measured in terms of the local $c$-efficiency 
	\[	
		\mathrm{eff}_c(\tau_0) =\frac{ \mathrm{aVar}(\hat{t}_{0.5};\tau^*) }{ \mathrm{aVar}(\hat{t}_{0.5};\tau_0)} = \frac{\mathbf{f}_2(\hat{t}_{0.5})^T \mathbf{M}_2(\tau^*)^{-1} \mathbf{f}_2(\hat{t}_{0.5})}{\mathbf{f}_2(\hat{t}_{0.5})^T \mathbf{M}_2(\tau_0)^{-1} \mathbf{f}_2(\hat{t}_{0.5})} = 98.70\%
	\]
	 for extrapolation at $t_{0.5}$ in the marginal mixed effects model and, hence, for estimation of the median failure time.
	This indicates that the adjusted design $\tau_0$ is highly efficient and can be favorably used as the time plan for conducting an accelerated degradation testing experiment.
\end{example}

Note that by Proposition~\ref{prop-extrapolation-t} the optimal design $\tau^*$ does not depend on the variance covariance structure $\BS\Sigma_{\gamma}$ of the random effects, but the efficiency of the adjusted design $\tau_0$ may be affected by the random effects, see equation (\ref{eq-var-t-extra-decompose}). 
Nevertheless, the $c$-efficiency $$\mathbf{f}_2(t_{0.5})^T \mathbf{M}_2^{(0)}(\tau^*)^{-1} \mathbf{f}_2(t_{0.5}) / \mathbf{f}_2(t_{0.5})^T \mathbf{M}_2^{(0)}(\tau_0)^{-1} \mathbf{f}_2(t_{0.5})$$ for extrapolation at $t_{0.5}$ in the fixed effect model provides a lower bound for the efficiency in estimating the median failure time.
For further approaches allowing for correlations depending on the distance of the measurement times, e.\,g.\ when measurements follow a random process, we refer to \citep{naether1985randomfields} and \citep{mueller2007spatialdesigns}.

\section{Cross-sectional time plans for destructive testing}
\label{sec-single-measurements}
In 
the case of destructive testing only $k = 1$ measurement will be available per unit. Hence, measurements have to be taken at different time points at different units in order to guarantee estimability of the parameters,
\begin{equation} 
	\label{eq-single-observation}
	Y_{i} = (\mathbf{f}_1(\mathbf{x}_{i}) \otimes \mathbf{f}_2(t_{i}))^T \BS{\beta} + \mathbf{f}_2(t_{i})^T\BS{\gamma}_i + \varepsilon_{i} \, ,
\end{equation}
$i = 1, ..., n$, where all other expressions have the same meaning as in the model equation~\ref{eq-general-model}.
We will restrict to the case of $k = 1$ measurements per unit in the remainder of this section.
In that case all measurements $Y_i$ are assumed to be independent. 
Denote by $\sigma^2(t) = \mathbf{f}_2(t)^T \BS{\Sigma}_\gamma \mathbf{f}_2(t) + \sigma^2_\varepsilon > 0$ 
the variance function for measurements at time $t$, $\mathrm{Var}(Y_i) = \sigma^2(t_i)$.
Also here an identifiability condition has to be imposed on the variance parameters $\BS\varsigma$ to distinguish between the variance $\sigma^2_\varepsilon$ of the measurement error and the variance $\sigma_1^2$ of the random intercept, see \cite{grasshoff2012optimal}.
The information matrix $\mathbf{M}_{\BS\beta}$ for the location parameter $\BS\beta$ can be written as
\begin{equation*}
	\label{eq-info-single-obs-exact}
		\mathbf{M}_{\BS\beta} = \sum_{i = 1}^n (\mathbf{f}_1(\mathbf{x}_i) \mathbf{f}_1(\mathbf{x}_i)^T) \otimes \left({\textstyle{\frac{1}{\sigma^2(t_i)}}} \mathbf{f}_2(t_i) \mathbf{f}_2(t_i)^T\right) 
\end{equation*}
according to the product-type structure of the model~(\ref{eq-single-observation}).
For estimating the median failure time $t_{0.5}$ the asymptotic variance is given by $\mathrm{aVar}(\widehat{t}_{0.5}) = c_0^2 \mathbf{c}^T \mathbf{M}_{\BS{\beta}}^{-1} \mathbf{c}$ as in Section~\ref{sec-failuretime}, where the vector $\mathbf{c}$ factorizes according to $\mathbf{c} = \mathbf{c}_1 \otimes \mathbf{c}_2$ into components $\mathbf{c}_1 = \mathbf{f}_1(\mathbf{x}_u)$ and $\mathbf{c}_2 = \mathbf{f}_2(t_{0.5})$ associated with the marginal models for the stress and the time variables, see (\ref{eq-avar-tau-alpha-decompose}).
The experimental settings $\mathbf{x}_i$ for the stress variable and $t_i$ for the time of measurement may be chosen independently from the standardized experimental regions $\mathcal{X}$ and $\mathcal{T}$, respectively. Here the approximate designs $\zeta$ are specified by $m$ mutually distinct combinations $(\mathbf{x}_i, t_i) \in \mathcal{X} \times \mathcal{T}$ with corresponding weights $\eta_i > 0$, $i = 1, ..., \nu$, $\sum_{i = 1}^\nu \eta_i = 1$.
The corresponding normalized information matrix is defined as 
\begin{equation*}
	\label{eq-info-single-obs-approx}
	\mathbf{M}(\zeta) = \sum_{i = 1}^\nu \eta_i (\mathbf{f}_1(\mathbf{x}_i) \mathbf{f}_1(\mathbf{x}_i)^T) \otimes (\tilde{\mathbf{f}}_2(t_i) \tilde{\mathbf{f}}_2(t_i)^T) \, , 
\end{equation*}
where $\tilde{\mathbf{f}}_2(t) = \mathbf{f}_2(t) / \sigma(t)$ is the weighted version of the marginal regression function for the time variable (standardized by the standard deviation for measurement at time $t$).
Then, in accordance with \citep{shat2021experimental}, a $c$-optimal design $\zeta^*$  for $\mathbf{c} = \mathbf{c}_1 \otimes \mathbf{c}_2$ can be obtained as the product-type design generated from the $c$-optimal designs $\xi^*$ for $\mathbf{c}_1$ and $\tau^*$ for $\mathbf{c}_2$ in the associated marginal models, respectively. 

\begin{theorem}
	\label{th-single-obs-prod-design}
	If $\xi^*$ is $c$-optimal for extrapolation at normal use condition $\mathbf{x}_u$ in the marginal model for the stress variables $\mathbf{x}$ and $\tau^*$  is $c$-optimal for extrapolation at the median failure time $t_{0.5}$ in the weighted marginal model with regression functions $\tilde{\mathbf f}_2(t)=\mathbf f_2(t)/\sigma(t)$ for the time variable $t$, then the design $\zeta^* = \xi^* \otimes \tau^*$ is optimal for estimating the median failure time $t_{0.5}$ in the combined model~(\ref{eq-single-observation}).
\end{theorem}

Note that for the second marginal model we utilize that $\tilde{\mathbf{f}}_2(t_{0.5})$ and $\mathbf{c}_2 = \mathbf{f}_2(t_{0.5})$ only differ by the factor $1 / \sigma(t_{0.5}) > 0$ which does not affect $c$-optimality. The $c$-optimal designs $\xi^*$ for extrapolation at $\mathbf{x}_u$ are the same as obtained in \citep{shat2021experimental}, and the $c$-optimal designs for extrapolation at $t_{0.5}$ can be obtained similarly by Elfving's theorem applied to the weighted regression functions $\tilde{\mathbf{f}}_2(t)$. 

The locally optimal designs for estimating the median failure time are influenced by both the location and the variance parameters $\BS\beta$ and $\BS\varsigma$.
Therefore we make a sensitivity analysis, how the optimal designs change with the parameters and how well the locally optimal designs perform under parameter misspecification.
For this we note first that the optimal marginal design $\xi^*$ for extrapolation at $\mathbf{x}_u$ is globally optimal and does not depend on the parameters.
In the particular case of straight line degradation paths, as in Example~\ref{ex-2}, the optimal marginal design $\tau^*$ is concentrated on the endpoints 0 and 1 of the standardized time interval, as long as the median $t_{0.5}$ is larger than the upper bound 1 of the interval. Then only the optimal weight $\pi^* = t_{0.5} \sigma(1) / (t_{0.5} \sigma(1) + (t_{0.5} - 1) \sigma(0))$ at $t=1$ depends on the location parameters $\BS\beta$ through $t_{0.5}$ and on the variance parameters $\BS\varsigma$ through the ratio $\sigma(1) / \sigma(0)$ of the standard deviations at the support points.

As in Example~\ref{ex-2}, the asymptotic eficiency
\[
	\mathrm{eff}_{\mathrm{aVar}}(\zeta; \BS\theta) = \frac{\mathrm{aVar}_{\BS\theta}(\hat{t}_{0.5}; \zeta^*_{\BS\theta})}{\mathrm{aVar}_{\BS\theta}(\hat{t}_{0.5}; \zeta)}
\]
of the design $\zeta$ for estimating $t_{0.5}$, factorizes, 
\[
	\mathrm{eff}_{\mathrm{aVar}}(\xi \otimes \tau; \BS\theta) = \mathrm{eff}_{c 1}(\xi) \cdot \mathrm{eff}_{c 2}(\tau; \BS\theta) \, ,
\]
in the case of product-type designs $\zeta = \xi \otimes \tau$ into the marginal efficiency $\mathrm{eff}_{c 1}(\xi)$ of $\xi$ for extrapolation at $\mathbf{x}_u$ and the marginal efficiency $\mathrm{eff}_{c 2}(\tau; \BS\theta)$ of $\tau$ for extrapolation at $t_{0.5}$. 
For a product-type design $\xi^*\otimes\tau$ with optimal first marginal design $\xi^*$, the asymptotic efficiency reduces to the $c$-efficiency of $\tau$ in the second marginal model,

\[
	\mathrm{eff}_{\mathrm{aVar}}(\xi^* \otimes \tau; \BS\theta) = \mathrm{eff}_{c 2}(\tau; \BS\theta) = \frac{\mathbf{c}_2(\BS\theta)^T \mathbf{M}_2(\tau^*_{\BS\theta}; \BS\theta)^{-1} \mathbf{c}_2(\BS\theta)}{\mathbf{c}_2(\BS\theta)^T \mathbf{M}_2(\tau; \BS\theta)^{-1} \mathbf{c}_2(\BS\theta)} \,.
\]
The vector $\mathbf{c}_2(\BS\theta) = \mathbf{f}_2(t_{0.5})$ depends on the location parameters $\BS\beta$ through $t_{0.5}$, but not on $\BS\varsigma$,
while the information matrices $\mathbf{M}_2(\tau; \BS\theta)$ are only affected by the variance parameters $\BS\varsigma$, but not by $\BS\beta$.
For straight line regression in the time variable (cf.\ Example~\ref{ex-2}), only the ratio $\sigma(1) / \sigma(0)$ of the standard deviations for measurements at the endpoints of the experimental time interval has an effect on the information matrix $\mathbf{M}_2(\tau; \BS\theta)$, when the design $\tau$ is supported by these endpoints, as is the case for the locally $c$-optimal design $\tau^*_{\BS\theta}$.
 In the following we revisit Example~\ref{ex-2} for obtaining an optimal design in regards to the current model of cross-sectional time plans. 

\begin{example*}[Example~\ref{ex-2} cont.]
 \label{ex-corss}
Under the current case of destructive testing, we provide in this example a $c$-optimal design of the time variable for the degradation model presented in Example~\ref{ex-2}. 
	For the stress variable, the $c$-optimal marginal design $\xi^*$ is derived in  \citep{shat2021experimental}. In particular, the optimal marginal design $\xi^*$ is concentrated on the endpoints $0$ and $1$ of the standardized design region with weight $w^*=0.05$ at $x=1$.
	For the time variable $t$, the $c$-optimal marginal design $\tau^*$ for extrapolation at $t_{0.5} > 1$ can be similarly obtained by Elfving's theorem (\citep{elfving1952}) applied to the vector of weighted regression functions $\tilde{\mathbf{f}}_2(t)= (1,t)^T / \sigma(t)$, where the variance function $\sigma^2(t)$ is defined here by  $\sigma^2(t) = \sigma_1^2+\rho\sigma_1\sigma_2 t+\sigma_2^2 t^2+\sigma_\varepsilon^2$.
	The shape of the Elfving set is exhibited in Figure~\ref{fig-elfving-set_tau}.
	\begin{figure}
		\centering
		\includegraphics[width=0.45\textwidth]{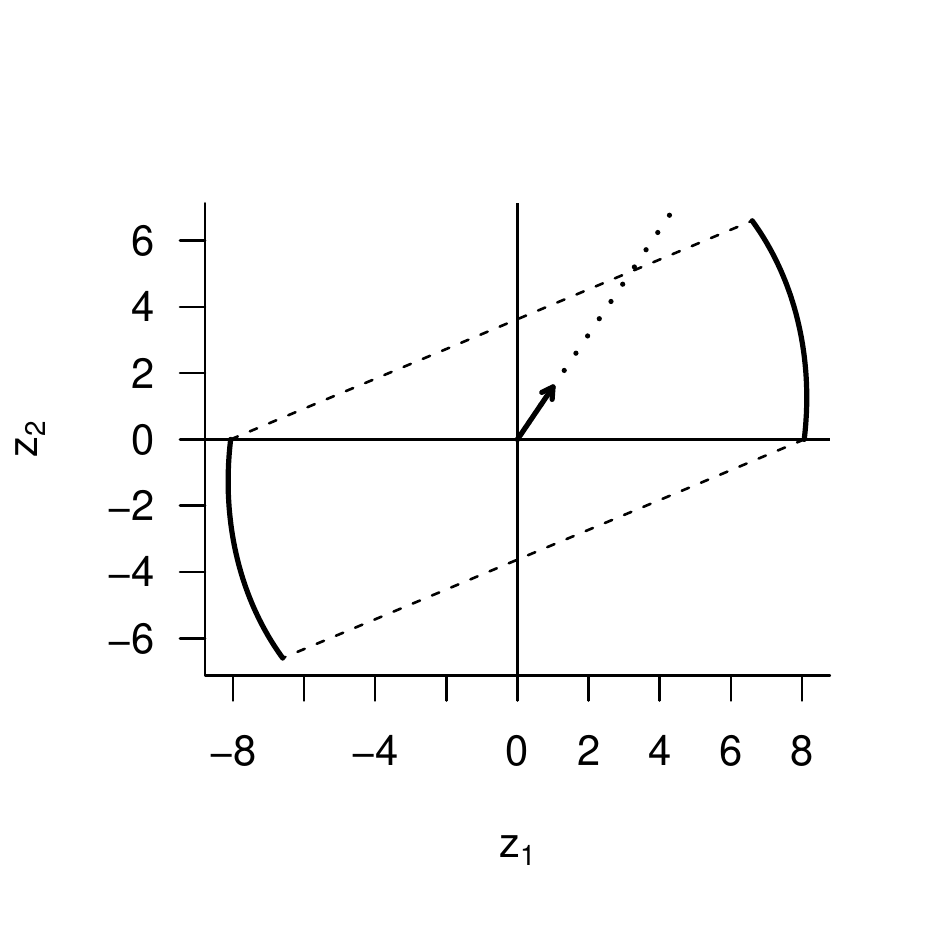}
		\caption{Elfving set for $t$ in Example~\ref{ex-2}: Induced design region (right solid line), negative image (left solid line), boundary (dashed lines), $\mathbf{c}_2 = (1, t_{0.5})^T $ (arrow) and corresponding ray (dotted line)}
		\label{fig-elfving-set_tau}
	\end{figure}
	The ray $\lambda \mathbf{f}_2(t_{0.5})= \lambda (1,t_{0.5})^T$ intersects the boundary of the Elfving set at the upper line segment connecting $\tilde{\mathbf{f}}_2(1) = (1, 1)^T / \sigma(1)$ and $ - \tilde{\mathbf{f}}_2(0) = (-1, 0)^T / \sigma(0)$ when $t_{0.5} > 1$.
	Hence, the $c$-optimal design $\tau^*$ is supported by the endpoints $0$ and $1$ of the standardized experimental time interval, and the optimal weights can be calculated by Elfving's theorem to $\pi^* = t_{0.5} \sigma(1) / (t_{0.5} \sigma(1) + (t_{0.5} - 1) \sigma(0))$ at $t = 1$ and $1 - \pi^* $ at $t = 0$.
	 
	By Theorem~\ref{th-single-obs-prod-design}, the design
	\[
		\zeta^* = \xi^* \otimes \tau^* = \left(
			\begin{array}{cccc}
				(0, 0) & (0, 1) & (1, 0) & (1, 1)
				\\
				(1 - w^*) (1 - \pi^*) & (1 - w^*) \pi^* & w^* (1 - \pi^*) & w^* \pi^*   
			\end{array}
		\right)
	\]
	is optimal for estimating the median failure time $t_{0.5}$.
	Under the nominal values of Table~\ref{tab-ex-intro-nominal-values} the optimal weights are $w^* = 0.05$ and $\pi^* = 0.77$.
	Then, the optimal design for estimating the median failure time is given by
	\[
		\zeta^* = \left(
		\begin{array}{cccc}
			(0, 0) & (0, 1) & (1, 0) & (1, 1)
			\\
			0.22 & 0.73 & 0.01 & 0.04 
		\end{array}
		\right) \, .
	\]

In Figure~\ref{fig-weight-t-ex1}, the optimal weight $\pi^* = t_{0.5} \sigma(1) / (t_{0.5} \sigma(1) + (t_{0.5} - 1) \sigma(0))$ is plotted  as a function of $t_{0.5}$ while the ratio $\sigma(1) / \sigma(0)$ is held fixed to $1.22$ induced by the nominal values of Table~\ref{tab-ex-intro-nominal-values}. 
\begin{figure}[!tbp]
  \centering
  \begin{minipage}[b]{0.396395\textwidth}
    \includegraphics[width=\textwidth]{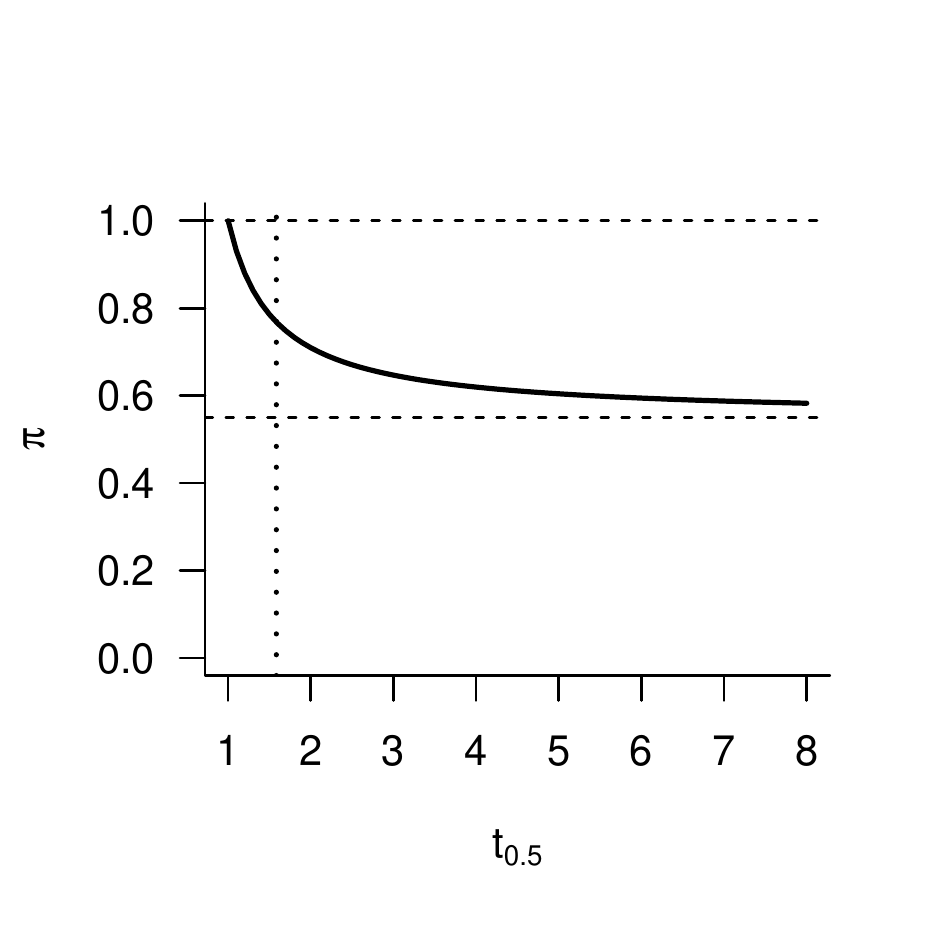}
    \caption{Optimal weights $\pi^*$ in dependence on $t_{0.5}$ for Example~\ref{ex-2}}
\label{fig-weight-t-ex1}
  \end{minipage}
  \hfill
  \begin{minipage}[b]{0.39639\textwidth}
    \includegraphics[width=\textwidth]{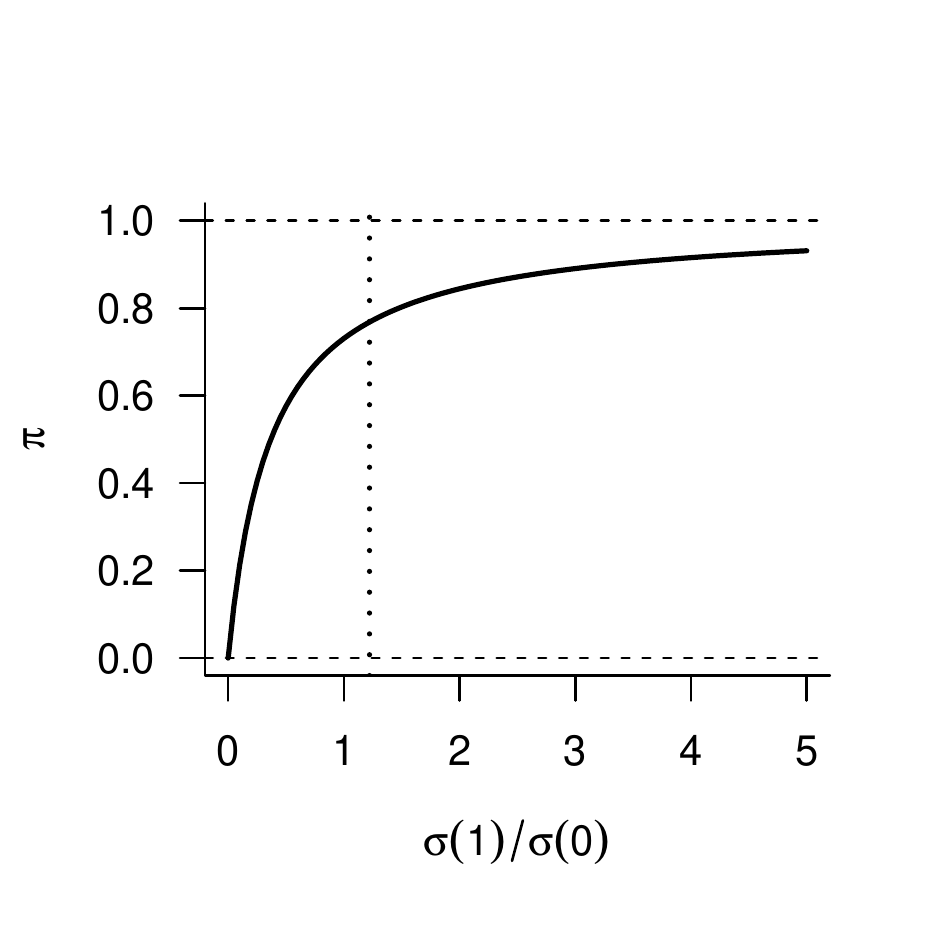}
   \caption{Optimal weights $\pi^*$ in dependence on $\sigma(1) / \sigma(0)$ for Example~\ref{ex-2}}
\label{fig-weight-sigma-ex1}
  \end{minipage}
\end{figure}
When $t_{0.5}$ increases, the optimal weight $\pi^*$ decreases from $1$ for $t_{0.5}$ close to the maximal experimental time $t_{\max} = 1$ to $\sigma(1) / (\sigma(1) + \sigma(0)) = 0.55$ for $t_{0.5} \to \infty$.
	In Figure~\ref{fig-weight-sigma-ex1} the optimal weight $\pi^*$ is plotted as a function of the ratio $\sigma(1) / \sigma(0)$ while $t_{0.5}$ is held fixed to $1.583$. On the other hand the optimal weight $\pi^*$ increases in the ratio $\sigma(1) / \sigma(0)$ from $0$ for the ratio close to $0$ to $1$ when the ratio tends to infinity.
	Those limiting values are indicated in Figures~\ref{fig-weight-t-ex1} and \ref{fig-weight-sigma-ex1} by horizontal dashed lines, and the nominal values for $t_{0.5}=1.583$ and $\sigma(1) / \sigma(0)=1.22$ are indicated by vertical dotted lines, respectively.

To assess the sensitivity of the optimal design with respect to parameter misspecification, we plot the efficiency of the locally optimal design $\zeta^* = \xi^* \otimes \tau^*$ in Figures~\ref{eff-t-ex1} and \ref{eff-sigma-ex1}.
For a comparison with standard marginal designs on the time scale, we also plot there the efficiency of the  designs $\xi^* \otimes \bar\tau_2$ (dashed line) and $\xi^* \otimes \bar\tau_6$ (dashed and dotted line), the marginal design $\bar\tau_k$ is uniform on $k$ equally spaced time points, i.\,e.\ $\bar\tau_2$ assigns weight $1 / 2$ to the endpoints $0$ and $1$ of the time interval, and $\tau_6$ assigns weight $1 / 6$ to the time points $t=0.0$, $0.2$, $0.4$, $0.6$, $0.8$, and $1.0$.

\begin{figure}[!tbp]
  \centering
  \begin{minipage}[b]{0.413996395\textwidth}
    \includegraphics[width=\textwidth]{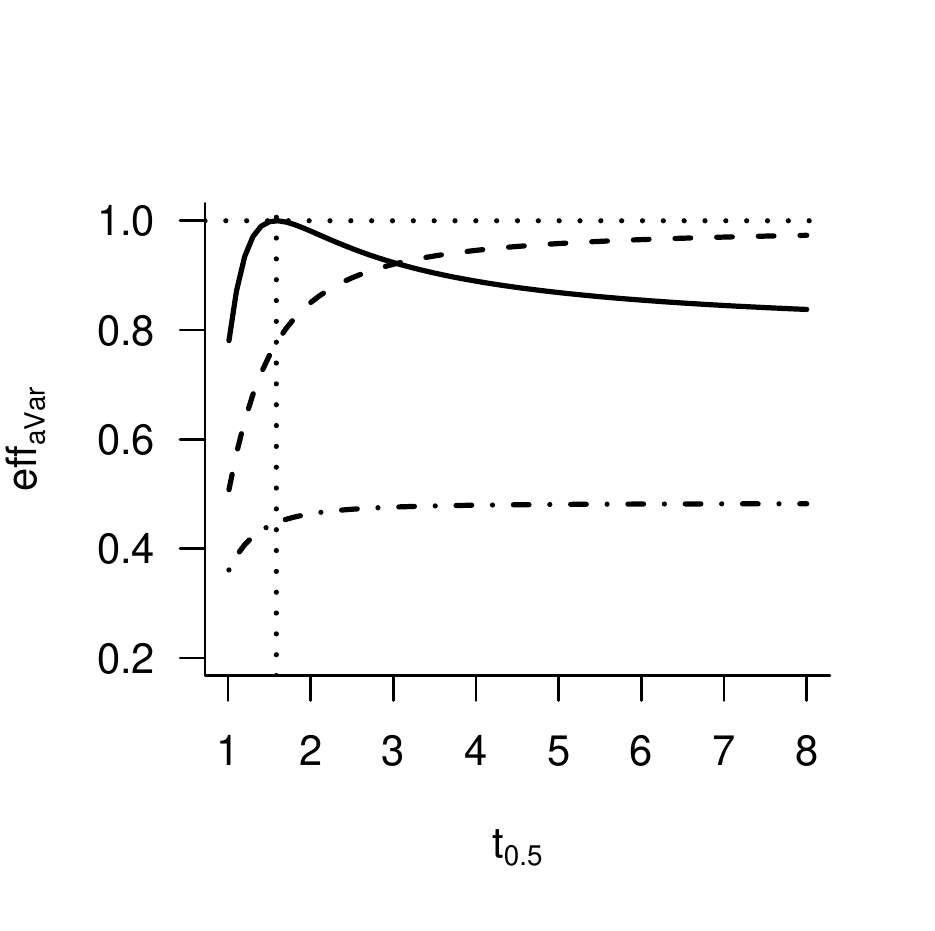}
    \caption{Efficiency of $\zeta^*=\xi^*\otimes\tau^*$ (solid line), $\xi^* \otimes \bar\tau_2$ (dashed line) and $\xi^* \otimes \bar\tau_6$ (dashed and dotted line)  in dependence on $t_{0.5}$ for Example~\ref{ex-2}}
\label{eff-t-ex1}
  \end{minipage}
  \hfill
  \begin{minipage}[b]{0.410396395\textwidth}
    \includegraphics[width=\textwidth]{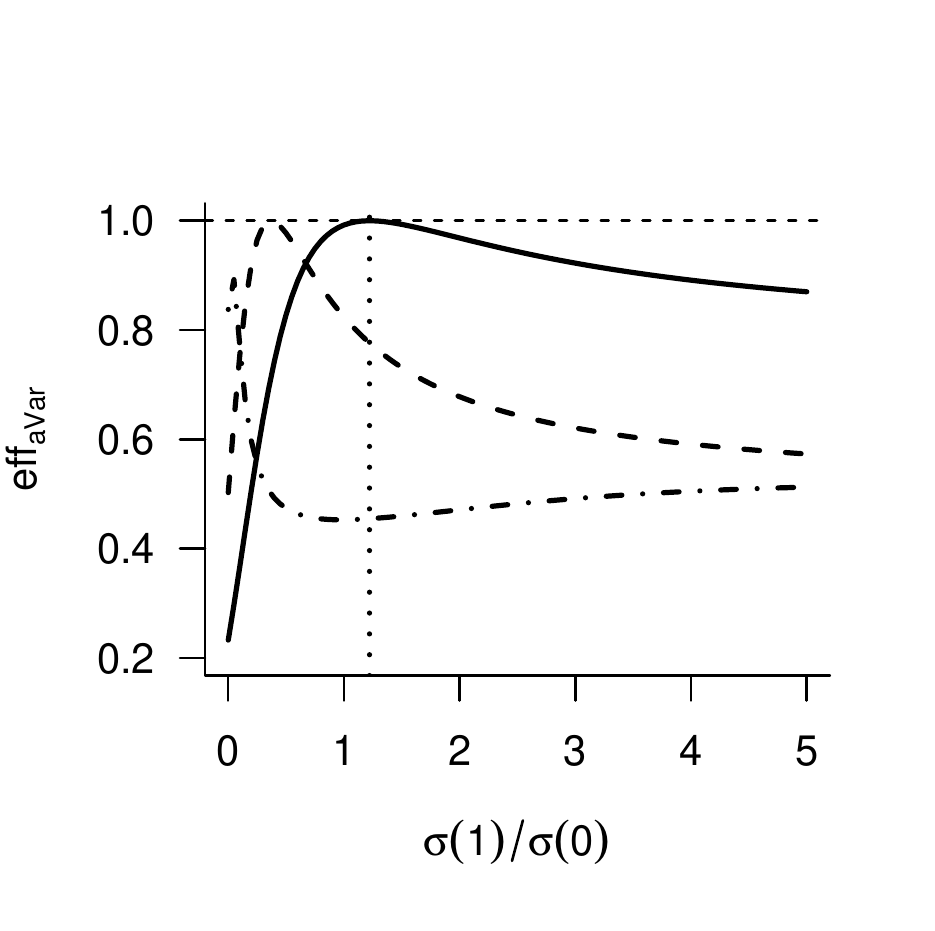}
   \caption{Efficiency of $\zeta^*=\xi^*\otimes\tau^*$ (solid line), $\xi^* \otimes \bar\tau_2$ (dashed line) and $\xi^* \otimes \bar\tau_6$ (dashed and dotted line)  in dependence on $\sigma(1) / \sigma(0)$ for Example~\ref{ex-2}}
\label{eff-sigma-ex1}
  \end{minipage}
\end{figure}

	In Figure~\ref{eff-t-ex1} the efficiency is displayed in dependence on the true value of the median failure time $t_{0.5}$ while the ratio $\sigma(1) / \sigma(0)$ of standard deviations is held fixed to the nominal value $1.22$. 
	In Figure~\ref{eff-sigma-ex1} the efficiency is shown in dependence on the true value of the ratio $\sigma(1) / \sigma(0)$ of standard deviations while the median failure time $t_{0.5}$ is held fixed to $1.583$ derived from the nominal values.
	Also here the nominal values for $t_{0.5}$ and $\sigma(1) / \sigma(0)$ are indicated by vertical dotted lines in the corresponding panel, respectively.

	The efficiency seems to be more sensitive with respect to deviations in the ratio of the standard deviations than in the median failure time $t_{0.5}$.
	However, commonly neither small values of $t_{0.5}$ nor small values of $\sigma(1) / \sigma(0)$ seem to be reasonable.
	In particular, we may expect that the variance $\sigma^2(1)$ at the end of the experimental time interval is larger than the variance $\sigma^2(0)$ at the initial time, $\sigma(1) / \sigma(0) \geq 1$.
	This is satisfied in the case of positive or no correlation, and it also holds for slight to moderate negative corrleations, $\rho \geq - \sigma_1 / (2\sigma_2)$.
	In total, the locally optimal design $\zeta^*=\xi^*\otimes\tau^*$ under the nominal values performs quite well over a wide range of parameters while the design $\xi^* \otimes \bar\tau_2$ which is concentrated on the same time points is only preferable for larger values of the median failure time $t_{0.5}$.
	The design $\xi^* \otimes \bar\tau_6$ with six equally spaced time points performs substantially worse throughout.
	It has to be noted for completeness that the efficiency of $\bar\tau_6$ depends on the variance parameters $\BS\varsigma$ not only through the ratio $\sigma(1) / \sigma(0)$ because measurements are also to be taken in the interior of the interval.
	To accomplish for this, the plot in Figure~\ref{eff-sigma-ex1} has been generated by fixing $\sigma_1^2 = \sigma_2^2 + \sigma_{\varepsilon}^2$ and letting $\rho$ vary.
	However, other choices of the variance parameters do not change much in the performance of the design. 
\end{example*}

\section{Discussion and conclusion}
\label{sec-discussion}

An elaborate assessment of the reliability related characteristics is of importance during the design stage of highly reliable systems.
The variability between units in accelerated degradation testing leads to a degradation function that can be described by a mixed-effects linear model.
This also results in a nondegenerate distribution of the failure time under normal use conditions such that it is desirable to estimate quantiles of this failure time distribution as a characteristic of the reliability of the product.
The purpose of optimal experimental design of the time plan is then to obtain the best settings for the time variable to obtain most accurate estimates for these quantities.

In the present model for accelerated degradation testing, it is assumed that in the experiment a cross-sectional design between units has to be chosen for the stress variable while for repeated measurements the time variable varies according to a longitudinal design within units.

Here we assumed a model with complete interactions between the time and the stress variables and random effects only associated with time but not with stress.
Then the cross-sectional design for the stress variables and the longitudinal design for the time variable can be optimized independently, and the resulting common optimal design can be generated as the cross-product of the optimal marginal designs for stress and time, respectively.
In particular, the same time plan for measurements can be used for all units in the test.

In contrast to that, in a model of destructive testing also the time variable has to be chosen cross-sectionally.
There the optimal choice of measurement times may also be affected by the variance covariance parameters of the random effects.
In both cases (longitudinal and cross-sectional time settings) the efficiency of the designs considered factorizes which facilitates to assess their performance when the nominal values for these parameters are misspecified at the design stage.

\renewcommand{\thesection}{\Alph{section}}
\setcounter{section}{0}

\section{Appendix}

\begin{proof}
	We prove the statement of the Lemma \ref{lem-repr-minv} by showing that multiplication of $\mathbf{M}$ with the right hand side $\mathbf{C} = (\mathbf{F}^T \BS\Sigma_\varepsilon^{-1} \mathbf{F})^{-1} + \BS\Sigma_\gamma$ results in the $k \times k$ identity matrix $\mathbf{I}_k$.
	For this note first that after premultiplication with $\mathbf{F}$  the right hand side can be expanded to 
	\[
		\mathbf{F} \mathbf{C} = \BS\Sigma_\varepsilon \BS\Sigma_\varepsilon^{-1} \mathbf{F} (\mathbf{F}^{T} \BS\Sigma_\varepsilon^{-1} \mathbf{F})^{-1} + \mathbf{F} \BS\Sigma_{\gamma} (\mathbf{F}^{T} \BS\Sigma_\varepsilon^{-1} \mathbf{F}) (\mathbf{F}^{T} \BS\Sigma_\varepsilon^{-1} \mathbf{F})^{-1} = \mathbf{V} \BS\Sigma_{\varepsilon}^{-1} \mathbf{F} (\mathbf{F}^{T} \BS\Sigma_{\varepsilon}^{-1} \mathbf{F})^{-1}.
	\] 
	Hence, by straightforward multiplication
	\[
		\mathbf{M} \mathbf{C} = \mathbf{F}^{T} \mathbf{V}^{-1} \mathbf{F} \mathbf{C} = \mathbf{F}^{T} \BS\Sigma_{\varepsilon}^{-1} \mathbf{F} (\mathbf{F}^{T} \BS\Sigma_{\varepsilon}^{-1} \mathbf{F})^{-1} = \mathbf{I}_k
	\]
	which proofs the lemma.
\end{proof}
\section*{Acknowledgement}
The work of the first author has been supported by the German Academic Exchange Service (DAAD) under grant no.~2017-18/ID-57299294.

\section*{References}
 \bibliographystyle{apa}
  \bibliography{ReferencesLMEM-timevariable}

\end{document}